\documentclass[twocolumn,showpacs,preprintnumbers,amsmath,amssymb]{revtex4-1}
\pdfoutput=1
\usepackage{graphicx}
\usepackage{rotating}
\usepackage{amssymb}
\usepackage{amsfonts}
\usepackage{amsmath}
\usepackage{color}

\begin{document}

\setcounter{MaxMatrixCols}{10}

\newtheorem{theorem}{Theorem}
\newtheorem{acknowledgement}[theorem]{Acknowledgement}
\newtheorem{algorithm}[theorem]{Algorithm}
\newtheorem{axiom}[theorem]{Axiom}
\newtheorem{claim}[theorem]{Claim}
\newtheorem{conclusion}[theorem]{Conclusion}
\newtheorem{condition}[theorem]{Condition}
\newtheorem{conjecture}[theorem]{Conjecture}
\newtheorem{corollary}[theorem]{Corollary}
\newtheorem{criterion}[theorem]{Criterion}
\newtheorem{definition}[theorem]{Definition}
\newtheorem{example}[theorem]{Example}
\newtheorem{exercise}[theorem]{Exercise}
\newtheorem{lemma}[theorem]{Lemma}
\newtheorem{notation}[theorem]{Notation}
\newtheorem{problem}[theorem]{Problem}
\newtheorem{proposition}[theorem]{Proposition}
\newtheorem{remark}[theorem]{Remark}
\newtheorem{solution}[theorem]{Solution}
\newtheorem{summary}[theorem]{Summary}
\newenvironment{proof}[1][Proof]{\noindent\textbf{#1.} }{\ \rule{0.5em}{0.5em}}

\renewcommand{\theequation}{\thesection.\arabic{equation}}

\newcommand{\re}{\mathop{\mathrm{Re}}}

\newcommand{\lb}{\label}
\newcommand{\be}{\begin{equation}}
\newcommand{\ee}{\end{equation}}
\newcommand{\bea}{\begin{eqnarray}}
\newcommand{\eea}{\end{eqnarray}}

\title{Varying constants entropic--$\Lambda$CDM cosmology}

\author{Mariusz P. D\c{a}browski}
\email{mpdabfz@wmf.univ.szczecin.pl}
\affiliation{\it Institute of Physics, University of Szczecin, Wielkopolska 15, 70-451 Szczecin, Poland,}
\affiliation{\it National Centre for Nuclear Research, Andrzeja So{\l}tana 7, 05-400 Otwock, Poland,}
\affiliation{\it Copernicus Center for Interdisciplinary Studies,
S{\l }awkowska 17, 31-016 Krak\'ow, Poland}

\author{H. Gohar}
\email{hunzaie@wmf.univ.szczecin.pl}
\affiliation{\it Institute of Physics, University of Szczecin, Wielkopolska 15, 70-451 Szczecin, Poland}

\author{Vincenzo Salzano}
\email{enzo@wmf.univ.szczecin.pl}
\affiliation{\it Institute of Physics, University of Szczecin, Wielkopolska 15, 70-451 Szczecin, Poland}

\date{\today}

\input epsf

\begin{abstract}

We formulate the basic framework of thermodynamical entropic force cosmology which allows variation of the gravitational constant $G$ and the speed of light $c$.
Three different approaches to the formulation of the field equations are presented. Some cosmological solutions for each framework are given and one of them is tested against combined observational data (supernovae, BAO, and CMB). From the fit of the data it is found that the Hawking temperature numerical coefficient $\gamma$ is two to  four orders of magnitude less than usually assumed on the geometrical ground theoretical value of $O(1)$ and that it is also compatible with zero. Besides, in the entropic scenario we observationally test that the fit of the data is allowed for the speed of light $c$ growing and the gravitational constant $G$ diminishing during the evolution of the universe. We also obtain a bound on the variation of $c$ to be $\Delta c/c \propto 10^{-5} >0$ which is at least one order of magnitude weaker than the quasar spectra observational bound.
\end{abstract}


\maketitle

\section{Introduction}
\label{intro}
\setcounter{equation}{0}

General Relativity is an established theory which explains the evolution of the universe on large scales \cite{EMM}. Although it is not complete because it
contains singularities, it explains the dynamics of the universe in a consistent way. Besides, the current phase of accelerated evolution of the universe has been discovered \cite{perlmutter,riess}. In order to obtain this accelerated expansion, one has to put an extra term, the cosmological constant $\Lambda $ or dark energy into the Einstein-Friedmann equations. Resulted $\Lambda $CDM models \cite{Peebles84, KofmanStar85,Ann86,Weinberg89} are consistent models to explain this accelerated expansion, but the observational value of $\Lambda $ is over 120 orders of magnitude smaller than the value calculated in quantum field theory, where it is interpreted as vacuum energy. This motivates cosmologists to look for alternative models which can explain the effect \cite{c2, c3}.

The relation between the Einstein's gravity and thermodynamics is a puzzle. In the seventies of the twentieth century, Bekenstein and Hawking \cite{BekensteinS,HawkingT}
derived the laws of black hole thermodynamics which emerged to have similar properties as in standard thermodynamics. Jacobson \cite{e} derived Einstein field equations from the first law of thermodynamics by assuming the proportionality of the entropy and the horizon area. A more extensive work in this direction was made by Verlinde and
Padmanabhan in Refs. \cite{f, f1,f2,f3}. Verlinde derived gravity as an entropic force, which originated in a system by the statistical tendency to increase its entropy. He assumed the holographic principle \cite{g}, which stated that the microscopic degrees of freedom can be represented holographically on the horizons, and this piece of information (or degrees of freedom) can be measured in terms of entropy. The approach got criticized on the base of neutron experiments though \cite{Archil}.

Recently, the entropic cosmology based on the notion of the entropic force was developed in series of papers \cite{komatsu,komatsu_proc,i1,i2,i3} and especially it was compared with supernovae data in Ref. \cite{easson}. However, supernovae test is not very strong and so the Ref. \cite{easson} got criticized on the base of galaxy formation problem (e.g \cite{Koivisto11,Basilakos12}). Basically, the idea of entropic cosmology is to add extra entropic force terms into the Friedmann equation and the acceleration equation. This force is supposed to be responsible for the current acceleration as well as for an early exponential expansion of the universe. It is pertinent to mention that the entropic cosmology suggested in these references assumes that gravity is still a fundamental force and that it includes extra driving force terms or boundary terms in the Einstein field equations. This is unlike Verlinde \cite{f}, who considers gravity as an entropic force, but not as a fundamental force (see also Refs. \cite{Cai05,Cai07,Cai08,Akbar06,Akbar071,Akbar072,Eling06,Hayward99,BakRay2000,Danielsson,Wang10}). All frameworks were discussed in detail in Ref. \cite{Visser11} by Visser. Entropic cosmology is also related to dynamical vacuum energy models which have been discussed and confronted with data in Refs. \cite{Basilakos09,Grande10,Grande11,Gomez141,Sola15}.

In this paper we expand entropic cosmology suggested in Refs. \cite{komatsu,komatsu_proc,i1,i2,i3,easson} for the theories with varying physical constants: the gravitational constant $G$ and the speed of light $c$. Although \cite{easson} is problematic in the context galaxy formation test, we use it as a starting point for further discussion. We discuss possible consequences of such variability onto the entropic force terms and
the boundary terms. As it has been known for the last fifteen years, varying constants cosmology \cite{VG,Vc} was proposed as an
alternative to inflationary cosmology, because it can to solve all the cosmological problems (horizon, flatness, and monopole). In the paper we try three different approaches to formulate the entropic cosmology with varying constants. In section \ref{field1} we present a consistent set of the field equations which describes varying constants entropic cosmology with general entropic force terms. In Section \ref{varthermo} we derive the continuity equation from the first law of thermodynamics and fit general entropic terms to the field equations derived in Section \ref{field1} using explitic definitions of Bekenstein entropy and Hawking temperature. We also discuss the constraints on the models which come from the second law of thermodynamics. In section \ref{modif1} we study single-fluid accelerating cosmological solutions to the field equations derived in Section \ref{varthermo}. In Section \ref{entropic} we derive the entropic force for varying constants, define appropriate entropic pressure, and modify the continuity and acceleration equations. We also determine Friedmann equation and give single-fluid accelerating cosmological solutions. In Section \ref{Enflow} we derive gravitational Einstein field equations using the heat flow through the horizon to which Bekenstein entropy and the Hawking temperature is assigned. Sections \ref{obspar} and \ref{data} are devoted to observationally testing the many-fluid entropic force models with varying constants. For this sake the data from supernovae, Baryon Acoustic Oscillations (BAO), and Cosmic Microwave Background (CMB) is used. In Section \ref{conclusions} we give our conclusions.

\section{Entropic force field equations and varying constants}
\label{field1}
\setcounter{equation}{0}

The main idea of our consideration is to follow Refs. \cite{easson,komatsu,komatsu_proc} (assuming homogeneous Friedmann geometry) and generalize field equations which contain the entropic force terms $f(t)$ and $g(t)$ onto the case of varying speed of light $c$ and varying Newton gravitational constant $G$ theories. It is easy to realize that the modified Einstein equations can be written down as follows
\begin{equation}
\left( \frac{\dot{a}}{a}\right) ^{2}=\frac{8\pi G(t)}{3}\rho - \frac{kc^2(t)}{a^2} + f\left(
t\right) ,
\label{22}
\end{equation}%
\begin{equation}
\frac{\ddot{a}}{a}=-\frac{4\pi G(t)}{3}\left[ \rho + \frac{3p}{c^{2}(t)}%
\right] +g(t).
\label{23}
\end{equation}%
In fact, the functions $f(t)$ and $g(t)$ in general play the role analogous to bulk viscosity (cf. Refs. [50-69] of the paper by Komatsu et. al. \cite{komatsu}) and this is why from (\ref{22}) and (\ref{23}) one obtains the modified continuity equation
\bea
\dot{\rho} &+& 3H\left[ \rho +\frac{p}{c^{2}(t)}\right] + \rho \frac{\dot{G(t)}}{G(t)} - 3 \frac{kc(t)\dot{c}(t)}{4\pi G(t)a^2(t)} \nonumber \\
&=& \frac{3H}{4\pi G(t)}\left[ g(t)-f(t)-\frac{\dot{f}(t)}{2H}\right] ,
\label{24}
\eea
which will further be used in our paper to various thermodynamical scenarios of the evolution of the universe. It is clear that (\ref{24}) has dissipative terms in full analogy to bulk viscosity models. However, if the functions $f(t)$ and $g(t)$ are equal and have the value of the $\Lambda$-term modified by varying speed of light $c(t)$ i.e.
\be
f(t) = g(t) = \frac{\Lambda c^2(t)}{3} ,
\ee
then they give modified varying $c$ and $G$ Einstein field equations with the continuity equation as \cite{JCAP15}
\begin{equation}
\label{conserLam}
\dot{\varrho}+3\frac{\dot{a}}{a}\left(\varrho+\frac{p}{c^2(t)} \right)+ \varrho\frac{\dot{G}(t)}{G(t)}= \frac{\left(3k-\Lambda a^2 \right)}{4\pi G(t)a^2}c(t)\dot{c}(t),
\end{equation}
which finally reduce to the standard $\Lambda$-CDM equations for $c$ and $G$ constant. Another point is the $f(t)$ and $g(t)$ terms can also be considered as time-dependent (dynamical) vacuum energy \cite{Basilakos09,Grande10,Grande11,Gomez141,Sola15}.

\section{Gravitational thermodynamics and varying constants}
\label{varthermo}
\setcounter{equation}{0}

In this section we start with basic thermodynamics in order to get entropic force varying constants field equations. We remind that the first law of thermodynamics
has been widely used to interlink different gravity theories with thermodynamics \cite{GW77,Cai05,Cai07,Cai08,Akbar06,Akbar071,Akbar072,Eling06}.
Defining the temperature and entropy on the cosmological horizons, one can use this law of thermodynamics for the whole universe
\begin{equation}
dE+pdV=TdS,  \label{a}
\end{equation}%
where $dE$, $dV$, and $dS$ describe changes in the internal energy $E$, the volume $V$, and the entropy $S$, while $T$ is the temperature, and $p$ is the pressure.
The volume of the universe contained in a sphere of the proper radius $r_{\ast} = a(t)r$ ($r$ is the comoving radius and $a(t)$ is the scale factor) is
\begin{equation}
V(t)=\frac{4}{3}\pi a^{3}r^{3}\text{ .}  \label{b}
\end{equation}%
We have
\be
\dot{V}(t)=3V(t)\frac{\dot{a}}{a}=3V(t)H(t)
\label{dervolume}
\ee
where dot represents the derivative with respect to time and the Hubble function is $H(t)=\dot{a}/a$. The internal energy $E$ and the energy density $\varepsilon (t)$ of the universe are related by
\begin{equation}
E(t)=\varepsilon (t)V(t),\hspace{0.3cm} \varepsilon (t)=\rho (t)c^{2}(t),
\label{c1}
\end{equation}%
where $\rho $ is the mass density of the universe.

Now we generalize the Hawking temperature $T$ \cite{HawkingT} and Bekenstein entropy $S$ \cite{BekensteinS} of the (time-dependent) Hubble horizon at $r \equiv r_h = r_h(t)$ onto the varying $c$ and $G$ theories as follows
\begin{eqnarray}
T&=&\frac{\gamma \hslash c(t)}{2\pi k_{B}r_{h}(t)} ,  \label{HT}\\
S&=&\frac{k_{B}}{%
4\hslash }\left[ \frac{c^{3}(t)A(t)}{G\left( t\right) }\right] . \label{BE}
\end{eqnarray}%
Here $A(t)=4\pi r_{h}^{2}(t)$ is the horizon area, $\hslash$ is the Planck constant, $k_{B}$ is the Boltzmann constant, and $\gamma $ is an
arbitrary, dimensionless, and non-negative theoretical parameter of the order of unity $O(1)$ which is usually taken to be $\frac{3}{2\pi }$, $\frac{3}{4\pi }$ or $\frac{1}{2}$
\cite{easson,komatsu,komatsu_proc}. In fact, $\gamma $ can be related to a corresponding screen or boundary of the universe to define the temperature and the entropy on that preferred screen. Here the screen will be the Hubble horizon i.e. the sphere of the radius $r_{h}$. Dividing (\ref{a}) by time differential $dt$, we have
\begin{equation}
\frac{dE}{dt}+p\frac{dV}{dt}=T\frac{dS}{dt},  \label{t1}
\end{equation}%
which after applying (\ref{b}) and (\ref{c1}) gives
\begin{equation}
\dot{E}+p\dot{V}=\left[ \dot{\rho}+2\frac{\dot{c}(t)}{c(t)}\rho +3\frac{\dot{%
a}}{a}\left( \rho +\frac{p}{c^{2}(t)}\right) \right] Vc^{2}(t).  \label{d3}
\end{equation}%
From (\ref{HT}) and (\ref{BE}) we have
\begin{equation}
T\dot{S}=\frac{\gamma c^{4}(t)}{2G(t)} r_h \left[ 3\frac{\dot{c}(t)}{c(t)}+2
\frac{\dot{r}_{h}}{r_{h}} - \frac{\dot{G}(t)}{G(t)}\right] .   \label{s1}
\end{equation}%
By using (\ref{t1}), (\ref{d3}), and (\ref{s1}) we get the modified continuity equation as follows
\bea
\dot{\rho} &+& 3H \left[ \rho +\frac{p}{c^{2}(t)}\right] = -2\frac{\dot{c}(t)}{c(t)}\rho
\label{ContinG} \\
&+& \frac{3\gamma H^{2}}{8\pi G(t)}\left[ \left( 5\frac{\dot{c}
(t)}{c(t)}-\frac{\dot{G}(t)}{G(t)}\right) -2\frac{\dot{H}}{H}\right]  \nonumber
\eea
where we have used the explicit definition of the Hubble horizon modified to varying speed of light models \cite{easson}
\begin{equation}
r_{h}(t) \equiv \frac{c(t)}{H(t)}.  \label{rh}
\end{equation}
If we introduced the non-zero spatial curvature $k=\pm 1$, then we would have to apply the entropy and the temperature of the apparent horizon which reads
\be
r_A=\frac{c(t)}{\sqrt{{H^{2}}+\frac{kc^{2}(t)}{a^{2}(t)}}}. \label{rA}
\ee
Simple calculations give that
\be
\frac{\dot{r}_A}{r_A} = - \frac{H r_A^2}{c^2} \left( \dot{H} - \frac{kc^2}{a^2} \right) + \frac{\dot{c}}{c} \left( 1 - \frac{k}{a^2} r_A^2 \right) ,
\ee
which for $k=0$ case reduces to
\be
\frac{\dot{r}_{h}}{r_{h}}=\frac{\dot{c}(t)}{c(t)}-\frac{\dot{H}}{H}.
\ee
In this section we restrict ourselves to $k=0$ case in order to get the general functions $f(t)$ and $g(t)$.

In order to constrain possible sets of varying constant models we can apply the second law of thermodynamics according to which the entropy of the universe remains
constant (adiabatic expansion) or increases (non-adiabatic expansion)
\be
\frac{dS}{dt}\geq 0 .
\ee
In fact, (\ref{s1}) gives the condition
\begin{equation}
3\frac{\dot{c}(t)}{c(t)}-\frac{\dot{G}(t)}{G(t)}\geq -2\frac{\dot{r}_{h}}{%
r_{h}}=-2\left( \frac{\dot{c}(t)}{c(t)}-\frac{\dot{H}}{H}\right) .
\label{225}
\end{equation}%
or
\begin{equation}
\label{bound}
5\frac{\dot{c}(t)}{c(t)}-\frac{\dot{G}(t)}{G(t)}\geq 2\frac{\dot{H}}{H} = 2 \left( \frac{\ddot{a}}{\dot{a}} - \frac{\dot{a}}{a} \right)
\end{equation}%
which for $\dot{c} = \dot{G} = 0$ just says that the Hubble horizon must increase $\dot{r}_h \geq 0$.
For $\dot{G}(t)=0,$ and by using (\ref{rh}) and (\ref{225}), we have
\begin{equation}
c\left( t\right) \geq b_{1}H^{\frac{2}{5}},
\label{2ndforc}
\end{equation}%
and for $\dot{c}(t)=0,$ we have%
\begin{equation}
G(t)\leq b_{2}H^{-2},
\label{2ndforG}
\end{equation}%
where $b_{1}$ and $b_{2}$ are constants.

\section{Gravitational thermodynamics -- cosmological solutions}
\label{modif1}
\setcounter{equation}{0}

Using the generalized continuity equation (\ref{ContinG}) one is able to fit the functions $f(t)$ and $g(t)$ from a general varying constants entropic force continuity equation (\ref{24}) as follows
\begin{eqnarray}
f(t) &=&\gamma H^{2}  \label{26} \\
g(t) &=&\gamma H^{2} + \frac{\gamma }{2}\left( 5\frac{\dot{c}(t)}{c(t)}-\frac{\dot{G}(t)}{
G(t)}\right) H \nonumber \\
&+& \frac{4\pi G(t)}{3H}\left( \frac{\dot{G}(t)}{G(t)}-2\frac{%
\dot{c}(t)}{c(t)}\right) \rho .  \label{27}
\end{eqnarray}%
Having given $f(t)$ and $g(t)$ one is able to write down the equations (\ref{22}) and (\ref{23}) as follows
\begin{equation}
\left( \frac{\dot{a}}{a}\right) ^{2}=\frac{8\pi G(t)}{3}\rho +\gamma H^{2},
\label{28}
\end{equation}
\begin{eqnarray}
\label{F29}
\frac{\ddot{a}}{a} &=& \gamma H^{2} - \frac{4\pi G(t)}{3}\left( \rho +\frac{3p}{c^{2}(t)}\right) \nonumber \\
 &+& \left( \frac{7\gamma -2}{2}\right) \frac{\dot{c}(t)}{c(t)}H+\left( \frac{1-2\gamma }{2}\right) \frac{\dot{G}(t)}{G(t)}H ,
\label{29}
\end{eqnarray}
which form a consistent set together with Eq. (\ref{ContinG}). While fitting the functions $f(t)$ and $g(t)$ we set $k=0$. If we were to investigate $k=\pm 1$ models then the the temperature $T$ (\ref{HT}) and the entropy $S$ (\ref{BE}) should be defined on the apparent horizon (\ref{rA}). A different choice of $f(t)$ and $g(t)$ which is consistent with (\ref{ContinG}) would be for example as follows
\begin{eqnarray}
f(t)& =&0 , \label{26a} \\
g(t) &=&\gamma  \dot{ H}+ \frac{\gamma }{2}\left( 5\frac{\dot{c}(t)}{c(t)}-\frac{\dot{G}(t)}{
G(t)}\right) H \nonumber \\
&+& \frac{4\pi G(t)}{3H}\left( \frac{\dot{G}(t)}{G(t)}-2\frac{%
\dot{c}(t)}{c(t)}\right) \rho .  \label{271}
\end{eqnarray}
However, both choices (\ref{26})-(\ref{27}) and (\ref{26a}) and (\ref{271}) do not allow for a constant term like the cosmological constant (unless one fine-tunes $H =$ const.) so that an alternative choice which fulfills this requirement would be
\begin{eqnarray}
f(t)& =&\gamma H^2+K_1 , \label{26aa} \\
g(t) &=&\gamma H^2+K_1 + \frac{\gamma }{2}\left( 5\frac{\dot{c}(t)}{c(t)}-\frac{\dot{G}(t)}{
G(t)}\right) H \nonumber \\
&+& \frac{4\pi G(t)}{3H}\left( \frac{\dot{G}(t)}{G(t)}-2\frac{%
\dot{c}(t)}{c(t)}\right) \rho ,  \label{271a}
\end{eqnarray}
where $K_1$ is a constant acting on the same footing as the cosmological $\Lambda$-term in standard $\Lambda$-CDM cosmology securing the model with respect to structure formation tests (cf. \cite{komatsu,Koivisto11,Basilakos12,Basilakos14,Gomez15}).

There is a full analogy of varying constants generalized equations (\ref{28}), (\ref{29}), and (\ref{ContinG}) with the entropic force equation given in \cite{easson} when one applies the specific ans\"atze for varying $c$ and $G$:
\be
\label{ansaetze}
c(t)=c_{o}a^{n} \hspace{0.5cm} {\rm and} \hspace{0.5cm} G(t)=G_{o}a^{q}
\ee
with $n, q=$ const. which gives $\dot{c}(t)/c(t)=nH$ or $\dot{G}(t)/G(t)=qH$. It is worth emphasizing that our ans\"atze should be $c(t)=c_{o}(a/a_0)^{n}$ and $G(t)=G_{o}(a/a_0)^{q}$ \cite{PLB14} but the standard approach nowadays picks up
$a_0 = 1$ \cite{amendola}.

We note that the application of the ans\"atze (\ref{ansaetze}) to the growing entropy requirement (\ref{bound}) gives the bound that
\be
n \geq -2/5 \hspace{0.8cm} {\rm and} \hspace{0.8cm} q \leq 2 .
\ee
As we shall see in Section \ref{data} (or Table I) these limits are in agreement with the observational values we have obtained. They also allow the Newtonian limit $c \to \infty$ ($n \to \infty$) or $G \to 0$ ($q \to - \infty$) \cite{okun}.

The cosmological solutions of the set of varying constants Eqs. (\ref{28}), (\ref{29}) and (\ref{ContinG}) are given below.

\subsection{$G$ varying models only: $G(t)=G_{o}a^{q};$ $q, G_{o} = const.$,
$\dot{c}(t)=0$.}

Defining the barotropic index equation of state parameter $w$ by using the barotropic equation of state, $p=w\rho c^{2}$ for varying $G=G_{o}a^{q}$, we
can integrate the continuity equation (\ref{ContinG}) to get
\begin{equation}
\rho =\rho _{0} a^{3\left( 1+w\right) } \left[ \left( \frac{G(t)}{G_o} \right) \left( \frac{H}{H_0} \right)^{2}\right] ^{\frac{\gamma }{\gamma -1}},  \label{n1}
\end{equation}%
where $\rho _{0}$ is a constant with the dimension of mass density, $G_o$ the gravitational constant, and $H_0$ the Hubble parameter. Using (\ref{28}) and (\ref{29}) and then multiplying (\ref{28}) by $\left( 1+3w\right)$ and (\ref{29}) by $2$ we get
\begin{equation}
\dot{H}=\left( \frac{1-2\gamma }{2}\right) \frac{\dot{G}}{G}H-\frac{3}{2}%
\left( 1+w\right) \left( 1-\gamma \right) H^{2}
\end{equation}%
or (using the fact that $\dot{G}/{G}=qH$, $\ddot{a}/a=\dot{H}+H^{2}$) one has
\begin{equation}
\dot{H}=-\bar{w}H^{2},  \label{n2}
\end{equation}%
where%
\begin{equation}
\bar{w}=\frac{1}{2}\left[ 3(w+1)\left( 1-\gamma \right) -(1-2\gamma )q\right].
\label{barw}
\end{equation}%
The Eq. (\ref{n2}) solves easily using a new variable $N=\ln a$ \cite{komatsu} i.e.
\begin{equation}
\left( \frac{dH}{da}\right) a=\frac{dH}{dN}=\left( \frac{dH}{dt}\right)
\frac{dt}{da}a=-\bar{w}H,
\end{equation}%
which integrates to give
\begin{equation}
H=H_0 a^{-\bar{w}},  \label{n5}
\end{equation}%
where $H_0$ is constant.
The solution of (\ref{n5}) is
\begin{equation}
a\left( t\right) =\bar{w}^{\frac{1}{\bar{w}}} \left[ H_0 \left( t-t_{0}\right) \right]^{\frac{1}{\bar{w}}},  \label{afinal}
\end{equation}%
where $t_{0}$ is constant. Bearing in mind the value of (\ref{barw}), we can easily conclude that
without entropic terms the solution (\ref{afinal}) corresponds to a standard
barotropic fluid Friedmann evolution $a(t)\propto (t-t_{0})^{(2/3(w+1))}$.
The scale factor for radiation, matter and vacuum (cosmological constant) dominated eras reads as
\begin{equation}
a(t)\propto \left\{
\begin{array}{c}
\left[H_0 (t-t_{0}) \right]^{\frac{2}{\left( 4-q\right) +2\gamma \left( q-2\right) }};w=\frac{1%
}{3},\text{ (radiation)} \\
\left[H_0 (t-t_{0}) \right]^{\frac{2}{\left( 3-q\right) +\left( 2q-3\right) \gamma }};w=0,%
\text{ (dust)} \\
\left[H_0 (t-t_{0})\right]^{\frac{2}{\left( 2\gamma -1\right) q}};w=-1.\text{ (vacuum)}%
\end{array}%
\right.   \label{aa}
\end{equation}%
The solution (\ref{aa}) shows that in varying $G$ entropic cosmology even dust ($w=0$) can drive acceleration of the universe provided
\begin{equation}
\left( 3-q\right) +\left( 2q-3\right) \gamma \leq 2~.
\end{equation}%
On the other hand, the solution which includes $\Lambda-$term $(w=-1)$ drives acceleration for $\left( 2\gamma -1\right) q\leq 2$.
There is an interesting check of these formulas for the case when one takes the Hawking temperature parameter $\gamma = 1$, in all three cases (radiation, dust, vacuum) the conditions for accelerated expansion fall into one relation $q \leq 2$. In fact, this limit is very special which can be seen from Eq. (\ref{28}) in which the terms involving $H^2$ cancel and lead to empty universe ($\varrho = 0$) so that it is no wonder that the acceleration does not depend on the barotropic index parameter $w$.
Finally, we conclude that in all these cases the entropic terms and the varying constants can play the role of dark energy.

One may also consider a more than one component models i.e. the models which allow matter, radiation as well as other cosmological fluids of negative pressure like the cosmological constant which give a turning point of the evolution compatible with current observational data (early-time deceleration and late-time acceleration). We will consider such models numerically in Section \ref{data}, where we test these models with observational data.

\subsection{$c$ varying models only: $c(t)=c_{o}a^{n};$ $c_0, n = const.$, $\dot{G}(t)=0$}

The solution of the continuity equation (\ref{ContinG}) for varying $c$ is
\begin{equation}
\rho =\rho _{0} a^{3\left( 1+w\right) } \left[ \left( \frac{c(t)}{c_0} \right)^{7\gamma -2} \left(\frac{H}{H_0} \right)^{-2\gamma } \right]^{\frac{1}{1-\gamma }},  \label{2a}
\end{equation}
where again $\rho _{0}$ is a constant with the dimension of mass density, $c_o$ the velocity of light, and $H_0$ the Hubble parameter. Applying (\ref{28}) and (\ref{29}), we
have
\begin{equation}
\dot{H}=\frac{7\gamma -2}{2}\frac{\dot{c}}{c}H-\frac{3}{2}\left( 1-\gamma
\right) \left( 1+w\right) H^{2},  \label{3a}
\end{equation}%
or
\begin{equation}
\dot{H}=-\tilde{w}H^{2},  \label{3aa}
\end{equation}%
where%
\begin{equation}
\tilde{w}=\frac{1}{2}\left[ 3(1+w)\left( 1-\gamma \right) -n\left( 7\gamma
-2\right) \right] .  \label{4a}
\end{equation}%
The solution of (\ref{3aa}) is
\begin{equation}
H=H_0 a^{-\tilde{w}},  \label{5a}
\end{equation}%
where $H_0$ is constant.
Finally, the solution of (\ref{5a}) for the scale factor gives
\begin{equation}
a(t)=\tilde{w}^{\frac{1}{\tilde{w}}} \left[H_0 \left( t-t_{0}\right) \right]^{\frac{1}{\tilde{w}}},
\end{equation}%
where $t_{0}$ is constant. For radiation, dust and vacuum we have, respectively
\begin{equation}
a(t)\propto \left\{
\begin{array}{c}
\left[H_0 (t-t_{0}) \right]^{\frac{2}{\left( 4+2n\right) -\left( 4+7n\right) \gamma }};w=\frac{
1}{3},\text{ (radiation)} \\
\left[H_0 (t-t_{0}) \right]^{\frac{2}{\left( 3+2n\right) -\left( 3+7n\right) \gamma }};w=0,
\text{ (dust)} \\
\left[H_0 (t-t_{0}) \right]^{\frac{2}{\left( 2-7\gamma \right) n}};w=-1.\text{ (vacuum)}
\end{array}
\right.
\end{equation}

For these three cases, one derives inflation provided
\begin{equation}
(4+2n)-(4+7n)\gamma\leq 2, \text{ (radiation)} \nonumber
\end{equation}
\begin{equation}
(3+2n)-(3+7n)\gamma\leq 2, \text{ (dust)} \nonumber
\end{equation}
\begin{equation}
(2-7\gamma)n\leq 2, \text{ (vacuum)} \nonumber
\end{equation}
and the entropic force terms play the role of dark energy which can be responsible for the current acceleration of the universe. As in the previous subsection, here also after taking the Hawking temperature parameter $\gamma = 1$, in all three cases (radiation, dust, vacuum) the conditions for accelerated expansion fall into one relation $n \geq - 2/5$, but this is also a special empty universe limit of Eq. (\ref{28}).

As in the previous subsection one may also consider a more than one component models -- the matter we deal with numerically in Section \ref{data}. We would like to emphasize again that here we have presented one-component solutions only while in Section \ref{obspar} we will be studying multi-component models which allow the transition from deceleration to acceleration.

\section{Entropic Pressure Modified Equations}
\label{entropic}
\setcounter{equation}{0}

In this section we start with the formal definition of the entropic force as given in \cite{komatsu,komatsu_proc,easson}. We assume that the temperature and entropy are given by (\ref{HT}) and (\ref{BE}) and use the definition of the entropic force
\begin{equation}
F=-T\frac{dS}{dr} . \label{o}
\end{equation}%
We calculate the entropic force on the horizon $r=r_{h}(t)$ by taking
\be
dS/dr_{h}=\dot{S}/\dot{r}_{h}
\ee
to obtain
\begin{equation}
F=-\frac{\gamma c^{4}(t)}{2G(t)}\left[ \frac{5\frac{\dot{c}(t)}{c(t)}-\frac{%
\dot{G}(t)}{G(t)}-2\frac{\dot{H}}{H}}{\frac{\dot{c}(t)}{c(t)}-\frac{\dot{H}}{%
H}}\right] .  \label{s}
\end{equation}%
For $\dot{c}=\dot{G}=0$ this formula reduces to the value obtained in Ref. \cite{komatsu}:
$F=\gamma (c^{4}/G)$ which is presumably the value of maximum tension in general relativity \cite{Gibbons2002,schiller,BG2014}. It has been shown in Ref. \cite{MaxTen} that (\ref{s}) may recover infinite tension thus violating the so-called Maximum Tension Principle \cite{Gibbons2002} in the framework of varying constants theories.

Now, we define the entropic pressure $p_{E}$, as the entropic force per unit area $A$, and use (\ref{rh}) to get
\begin{equation}
p_{E}=-\frac{\gamma c^{2}(t)H^2}{8\pi G(t)}\left[ \frac{5\frac{\dot{c}(t)
}{c(t)}-\frac{\dot{G}(t)}{G(t)} - 2\frac{\dot{H}}{H}}{\frac{\dot{c}(t)}{%
c(t)}-\frac{\dot{H}}{H}}\right] .  \label{u}
\end{equation}%
Out of the set of initial equations (\ref{22})-(\ref{24}) only two of them are independent. On the other hand, only (\ref{23}) (acceleration equation) and (\ref{24}) (continuity equation) contain the pressure. This is why while having (\ref{u}) we will define the effective pressure
\be
p_{eff}=p+p_{E} \label{peff}
\ee
and then write down the continuity equation (\ref{24}) as
\begin{equation}
\dot{\rho}+3H\left( \rho +\frac{p_{eff}}{c^{2}(t)}\right)+\frac{\dot{G}(t)}{G(t)}
\rho =0,  \label{v}
\end{equation}
or
\bea
\dot{\rho}&+& 3H\left( \rho +\frac{p}{c^{2}(t)}\right) +\frac{\dot{G}(t)}{G(t)}
\rho \\
&=& \frac{3\gamma H^3}{8\pi G(t)}\left[ \frac{5\frac{\dot{c}(t)}{c(t)}-
\frac{\dot{G}(t)}{G(t)} - 2\frac{\dot{H}}{H}}{\frac{\dot{c}(t)}{c(t)}-%
\frac{\dot{H}}{H}}\right] , \nonumber  \label{w}
\eea
and the acceleration equation (\ref{23}) as
\begin{equation}
\frac{\ddot{a}}{a}=-\frac{4\pi G(t)}{3}\left( \rho +\frac{3p_{eff}}{c^{2}(t)}%
\right)   \label{x}
\end{equation}%
or%
\begin{equation}
\frac{\ddot{a}}{a}=-\frac{4\pi G(t)}{3}\left( \rho +\frac{3p}{c^{2}(t)}%
\right) +\frac{\gamma H^2}{2}\left[ \frac{5\frac{\dot{c}(t)}{c(t)}-\frac{
\dot{G}(t)}{G(t)} - 2 \frac{\dot{H}}{H}}{\frac{\dot{c}(t)}{c(t)}-\frac{\dot{%
H}}{H}}\right]   \label{y}
\end{equation}
In order to solve the continuity equation (\ref{w}) we have to put $f(t)=0$ in the Friedmann equation. Alternatively, we see by comparing (\ref{w}) and (\ref{24}) for $k=0$, that we need to put $f(t)=0$. We then obtain the simplest form of the Friedmann equation to use
\begin{equation}
\left( \frac{\dot{a}}{a}\right) ^{2}=\frac{8\pi G(t)}{3}\rho,
\label{28a}
\end{equation}

By using (\ref{y}) and (\ref{28a}), we get for varying $c(t)=c_{0}a^{n}$ and $G(t)=G_{0}a^{q}$:
\begin{equation}
\left(\dot{H}\right)^{2}-\left(\frac{B_1+2n}{2}\right)\dot{H}H^{2}-\left(\frac{nB_2+q\gamma}{2}\right)H^4=0,
\label{Hdot2}
\end{equation}
where
\begin{equation}
B_1=-3(1+w)+2\gamma
\end{equation}
\begin{equation}
B_2=3(1+w)-5\gamma
\end{equation}

The cosmological solutions are obtained below. We consider two cases.

\subsection{{\bf $G$ varying models only:} $\dot{G}(t)\neq 0$ and $\dot{c}(t)=0$; $q \neq 0, n=0$.}

The Eq. (\ref{Hdot2}) reduces to
\begin{equation}
\left(\dot{H}\right)^{2}-\left(\frac{B_1}{2}\right)\dot{H}H^{2}-\left(\frac{q\gamma}{2}\right)H^4=0,
\end{equation}
or we can write
\begin{equation}
\dot{H}-\frac{B_{1}H^{2}}{4}=\pm\sqrt{\left(\frac{q\gamma}{2}+\frac{B^{2}_{1}}{16}\right)} H^{2}
\end{equation}
or
\begin{equation}
\dot{H}=-WH^2, \label{WH2}
\end{equation}
where
\begin{equation}
W=\mp\sqrt{\left(\frac{q\gamma}{2}+\frac{B^{2}_{1}}{16}\right)}+\frac{B_1}{4}.
\end{equation}
Solving (\ref{WH2}) for the Hubble parameter, we have
\begin{equation}
H=H_0 a^{-W} , \label{HKW}
\end{equation}
where $H_0$ is a constant of integration. Solving (\ref{HKW}) for the scale factor $a(t)$, one gets
\begin{equation}
a(t)=W^{\frac{1}{W}}\left[H_0 (t-t_{0})\right]^{\frac{1}{W}}.
\end{equation}

\subsection{{\bf $c$ varying models only:} $\dot{G}(t)=0$ and $\dot{c}(t)\neq 0$; $q=0, n \neq 0$}

From (\ref{Hdot2}) we obtain
\begin{equation}
\left(\dot{H}\right)^{2}-\left(\frac{B_1+2n}{2}\right)\dot{H}H^{2}-\left(\frac{nB_2}{2}\right)H^4=0,
\end{equation}
Following the same procedure as in the subsection A, one can find the Hubble parameter and the scale factor for varying $c$ as:
\begin{equation}
H=H_0 a^{-X}
\end{equation}
and
\begin{equation}
a(t)=X^{\frac{1}{X}}\left[H_0 (t-t_{0})\right]^{\frac{1}{X}},
\end{equation}
where, $H_0$ and $t_{0}$ are real constants and $X$ is given by
\begin{equation}
X=-\left(\pm\sqrt{\left(\frac{nB_{2}}{2}+\frac{(B_{1}+2n)^{2}}{16}\right)}+\frac{B_1+2n}{4}\right).
\end{equation}
Both of the above cases have the same non-varying constants limit ($n \to 0$ or $G \to 0$) of $W = B_1/2$.

\section{Gravitational thermodynamics - horizon heat flow}
\label{Enflow}
\setcounter{equation}{0}

In this section we use yet another approach to derive entropic cosmology which is based on the application of the idea that one can get gravitational Einstein field equations using the heat flow through the horizon to which Bekenstein entropy (\ref{BE}) and Hawking temperature (\ref{HT}) (with $\gamma = 1$) are assigned.

The heat flow $dQ$ out through the horizon is given by the change of energy $dE$ inside the apparent horizon and relates to the flow of entropy $TdS$ as follows \cite{Danielsson,BakRay2000,Hayward99}
\be
\label{dQ}
dQ = TdS = - dE.
\ee
If the matter inside the horizon has the form of a perfect fluid and $c$ is not varying, then the heat flow through the horizon over the period of time $dt$ is \cite{BakRay2000}
\be
\frac{dQ}{dt} = T \frac{dS}{dt} = A(\varrho + \frac{p}{c^2}) = 4\pi r_A^2 (\varrho + \frac{p}{c^2})
\ee
However, in our case $c$ is varying in time and we have to take this into account while calculating the flow so that bearing in mind that the mass element is $dM$ we
have the energy through the horizon as
\begin{equation}
-dE =c^{2}dM+2Mcdc+pdV.  \label{energyflow2}
\end{equation}
The mass element flow is
\be
dM = A (v dt) \varrho = dV \varrho ,
\label{dM}
\ee
where $v dt = s$ is the distance travelled by the fluid element, $v$ is the velocity of the volume element, and $dV$ is the volume element. The velocity of a fluid element can be related to the Hubble law of expansion
\be
v = H r_A
\label{HrA}
\ee
so that (\ref{dM}) can be written down as
\be
dM = A H r_A \varrho dt .
\label{AHrA}
\ee
We assume that the speed of light is the function of the volume through the scale factor i.e. $c = c(V)$ and since $a \propto V^{1/3}$, then $c = c(a)$ \cite{youm}. We have
\be
\frac{dc}{dV} = \frac{1}{3} \frac{1}{V^{2/3}} \frac{dc}{da}
\ee
and besides by putting $M = V \varrho$ in (\ref{energyflow2}) we get
\be
-dE = c^2 dV \left( \varrho + \frac{p}{c^2} + \frac{2}{3} \varrho \frac{a}{c} \frac{dc}{da} \right) .
\label{dEfin}
\ee
Using (\ref{AHrA}), (\ref{dEfin}) and (\ref{s1}) (replacing $r_h$ by $r_A$) one has from (\ref{dQ})
\be
4 \pi r_A^2 H \left(\varrho + \frac{p}{c^2}+ \frac{2}{3} \varrho \frac{a}{c} \frac{dc}{da} \right) = \frac{c^2}{2G} \left( 3\frac{\dot{c}}{c} + 2 \frac{\dot{r}_A}{r_A} - \frac{\dot{G}}{G} \right) ,
\ee
or after explicitly using (\ref{rA}) we get a generalized acceleration equation
\bea
\dot{H}&=&-4\pi G(\rho +\frac{p}{c^{2}})+\frac{1}{2}\left( 5\frac{\dot{c}}{c}- \frac{\dot{G}}{G}\right) H \nonumber \\
&-& \frac{8\pi G}{3} \frac{\dot{c}}{c} \frac{\rho}{H} + \frac{1}{2} \frac{kc^2}{a^2 H} \left( \frac{\dot{c}}{c} - \frac{\dot{G}}{G} + 2H \right)  ,
\label{acceleq}
\eea
which for $\dot{c} = \dot{G} = 0$ gives the Eq. (A6) of Ref. \cite{easson}.

In order to get the Friedmann equation, we have to use the continuity equation (\ref{h}) but for adiabatic expansion ($dS=0$) to obtain
\begin{equation}
H\left( \rho +\frac{p}{c^{2}}\right) =-\frac{\dot{\rho}}{3}-\frac{2}{3}\frac{\dot{c}}{c}\rho.   \label{conteq}
\end{equation}
By using Eq. (\ref{acceleq}) in (\ref{conteq}), we have
\bea
H\dot{H}&=&\frac{4\pi G(t)}{3}\dot{\rho}+\frac{1}{2}\left( 5\frac{\dot{c}}{c}-\frac{\dot{G}}{G}\right) H^{2} \nonumber \\
&+& \frac{1}{2} \frac{kc^2}{a^2 H} \left( \frac{\dot{c}}{c} - \frac{\dot{G}}{G} + 2H \right).
\label{new}
\eea
After integrating (\ref{new}) one obtains a generalized Friedmann equation
\bea
H^{2}&=&\frac{8\pi }{3}\int G(t)\dot{\rho}dt + \int \left( 5\frac{\dot{c}}{c}- \frac{\dot{G}}{G}\right) H^{2}dt \nonumber \\
&+& \frac{1}{2} \int\frac{kc^2}{a^2 H} \left( \frac{\dot{c}}{c} - \frac{\dot{G}}{G} + 2H \right)
\label{friedeq}
\eea
For $k=0$ ($r_A \to r_h = c/H$) by taking the ansatz of the form
\be
c(t)=c_0 [H(t)]^m ,
\label{Bansatz}
\ee
$c_0=$ const., $m=$ const. (or $c(t) = c_0 (H/H_0)^m$, $H_0=$ const.; similar ansatz $c(t) = \dot{a}(t)$ was used in Ref. \cite{buchalter}), for varying $c$ only (i.e. for $\dot{G} = 0$), we have the following equations
\begin{equation}
H^{2}=\frac{8\pi G}{3}\rho+\frac{5m}{2}H^2 + K,
\label{Ban1}
\end{equation}%
\begin{equation}
\frac{\ddot{a}}{a}=-\frac{4\pi G}{3}(\rho +\frac{3p}{c^{2}})+\frac{5m}{2}H^2+(\frac{3m}{2}+\frac{5m^2}{2})\dot H+K+mK\frac{\dot H}{H^2},
\label{Ban2}
\end{equation}%
\begin{equation}
\dot{\rho}+3H\left( \rho +\frac{p}{c^{2}(t)}\right) +2m\frac{\dot{H}}{H}\rho =0,
\label{Ban3}
\end{equation}
where solely $K$ is the constant of integration which can be interpreted as the cosmological constant $\Lambda c^2/3$ \cite{Danielsson} provided $m=0$ (cf. the discussion in Section \ref{field1} and the formula (\ref{conserLam})). For small $m$ one may expand $c(t)$ given by (\ref{Bansatz}) in Taylor series
\bea
c(t) &=& c_0 \left[H(t)\right]^m \\
&=& c_0 \left[ 1 + m \ln{H(t)} + \frac{m^2}{2} \left( \ln{H(t)} \right)^2 + \ldots \right] \nonumber
\eea
and use numerical procedures to calculate the consequences of variability of the speed of light but we keep this beyond the scope of the paper.

Our set of equations (\ref{Ban1})-(\ref{Ban3}) contains two effects: the entropic force
contribution $K$ as well as many new terms related to variability of $c$ (all the terms which involve the parameter $m$). In fact, there are as many as four such latter terms in the equation (\ref{Ban2}) (including a cross-term with $K$) and each of them may play the role in accelerating the universe instead of $K$-term.

For $K=0$ case one can easily solve for the Hubble parameter $H$ and the scale factor $a$ for varying $c$ models as
\begin{equation}
H=H_0 a^{-\frac{3(1+w)}{2(1+m)}},
\end{equation}
 and
\begin{equation}
a(t)\propto \left[H_0 (t-t_0)\right]^{\frac{2(1+m)}{3(1+w)}}.
\end{equation}
where $H_0$ is the constant of integration. Besides, the continuity equation solves by
\begin{equation}
\rho=\rho_0 a^{-\frac{3(1+w)}{1+m}},
\end{equation}
where $\rho_0$ is a constant with the dimension of mass density. The solutions for $K \neq 0$ can be found numerically, but we do not present them here.

\section{Observational Parameters}
\label{obspar}

In this section we will try to give some more quantitative information about our approach, by applying our model to observational data. We will leave the single-fluid approach we have considered in past section, to move to the more realistic case of a multi-fluid scenario. We will take into account the components which make up the total mass density $\rho$, i.e., radiation $\rho_r$, matter $\rho_m$, and some unknown vacuum energy component $\rho_v$ (which can also be the cosmological constant). We take the model with $f(t)$ and $g(t)$ given by (\ref{26}) and (\ref{27}) which do not contain the constant $K_1$-term as in (\ref{26aa}) and (\ref{271a}). However, we will get this constant term effectively as the energy density of vacuum $\rho_{\Lambda} = (\Lambda c^2)/(8\pi G)$. With these assumptions we can write the continuity equation (\ref{ContinG}) as
\bea
\sum_{i} \dot{\rho_{i}} &+& 3H \left[ \sum_{i}\rho_{i} +\frac{\sum_{i} p_{i}}{c^{2}(t)}\right] = -2\frac{\dot{c}}{c} \sum_{i}\rho_{i}
\label{h} \\
&+& \frac{\gamma}{1-\gamma} \sum_{i} \rho_{i} \left[ \left( 5\frac{\dot{c}
(t)}{c(t)}-\frac{\dot{G}(t)}{G(t)}\right) -2\frac{\dot{H}}{H}\right]  \nonumber
\eea
where summation on $i$ runs for radiation, matter and dark energy. From (\ref{h}) one can easily check that one can separate the contribution of the three fluids obtaining three separate continuity equations; if we introduce the equation of state parameter $w_i$ from $p_i = w_i \; \rho_i$, then the continuity equations have the form:
\bea \label{continuity_new}
\dot{\rho_{i}} &+& 3H \rho_{i} \left( 1+ w_{i} \right) = -2\frac{\dot{c}}{c} \rho_{i}
\label{rhomulti} \\
&+& \frac{\gamma}{1-\gamma} \rho_{i} \left[ \left( 5\frac{\dot{c}
(t)}{c(t)}-\frac{\dot{G}(t)}{G(t)}\right) -2\frac{\dot{H}}{H}\right]  \nonumber \; ,
\eea
where, of course, $w_{i} = 0$ for matter, $w_{i} = 1/3$ for radiation, and $w_{i} = -1$ for vacuum. From them, it is easy to check that, on one hand, no interaction term is present, in the way of exchanging energy among the fluids; but, on the other, entropic forces and the varying constants influence the behaviour of the fluids, by the same amount and separately. Thus, for each of them, a separate continuity equation holds, and we never have any violation of the mass-energy conservation law.

The solution for each fluid from (\ref{continuity_new}) can be easily found; once we use our ansatze, $c = c_{0} a^n$ and $G = G_{0} a^q$, we have:
\begin{equation}
\rho_{i} = \frac{\rho_{0}}{H_{0}^{\frac{2\gamma}{1-\gamma}}} H^{\frac{2\gamma}{1-\gamma}} a^{f^{X}_{i}(\gamma,n,q)} \;
\end{equation}
where, as usual, $H$ is the Hubble function, $H_{0}$ the Hubble constant, $a$ the scale factor, and $f_{i}(\gamma,n,q)$ are general functions obtained by solving (\ref{continuity_new}). When considering only a varying $c$, these functions are:
\begin{equation}
f^{c}_{i}(\gamma,n) = - 3 \left[ 1 + w_{i} + \frac{n(2-7\gamma)}{3(1-\gamma)}\right]\; ,
\end{equation}
while for a varying $G$ we have:
\begin{equation}
f^{G}_{i}(\gamma,q) = - 3 \left[ 1 + w_{i} + \frac{q \gamma}{3(1-\gamma)}\right]\; .
\end{equation}
We can note down that the main changes to the equation of state parameters come for the varying constant assumptions: in the limit of $n \rightarrow 0$ and $q \rightarrow 0$, we recover the usual behaviours, $a^{-3}$ for matter, $a^{-4}$ for radiation, and constancy for the vacuum. But still we have some dynamical effects on the densities from the entropic forces, through the $H^{\frac{2\gamma}{1-\gamma}}$ term. Thus, even in the case of no-varying constant, the entropic forces make the vacuum dynamical.

Starting from the Friedmann equation (\ref{28}), after some simple algebra, we can write the Hubble function $H$, which we explicitly need for observational fitting. In the case of varying $c$ it will be
\begin{equation}
E^{2}=\left(\frac{H}{H_{0 }}\right)^{2}= \left[ \sum_{i} \frac{\Omega_{i,0}}{1-\gamma} a^{f^{c}_{i}}(\gamma,n) \right]^{\frac{\gamma-1}{2\gamma-1}}
\end{equation}
while for varying $G$ it will be
\begin{equation}
E^{2}=\left(\frac{H}{H_{0 }}\right)^{2}= \left[ a^{q} \sum_{i} \frac{\Omega_{i,0}}{1-\gamma} a^{f^{G}_{i}}(\gamma,q) \right]^{\frac{\gamma-1}{2\gamma-1}}
\end{equation}
We have defined the dimensionless density parameters as
\begin{equation}
\Omega_{i}=\frac{8\pi G_{0}\rho_{i,0}}{3H^{2}_{0}},
\end{equation}
where, $G_{0}$ is the current value of Newton's gravitational constant. Finally, in order to check if our model (\ref{h}) allows a transition from deceleration to acceleration during the evolution of the universe at some redshift $z$ in a similar way to a ``pure'' $\Lambda$CDM model, we have looked at the deceleration parameter, defined as:
\begin{equation}
\label{q(z)}
q(z) = \frac{(1+z)}{2\,H^{2}(z)}\frac{dH^{2}(z)}{dz} - 1\; ,
\end{equation}
where the cosmological redshift is given by $1+z = 1/a$.

\begin{figure*}[htbp!]
\begin{minipage}{\textwidth}
 \centering
  \includegraphics[width=0.45\textwidth]{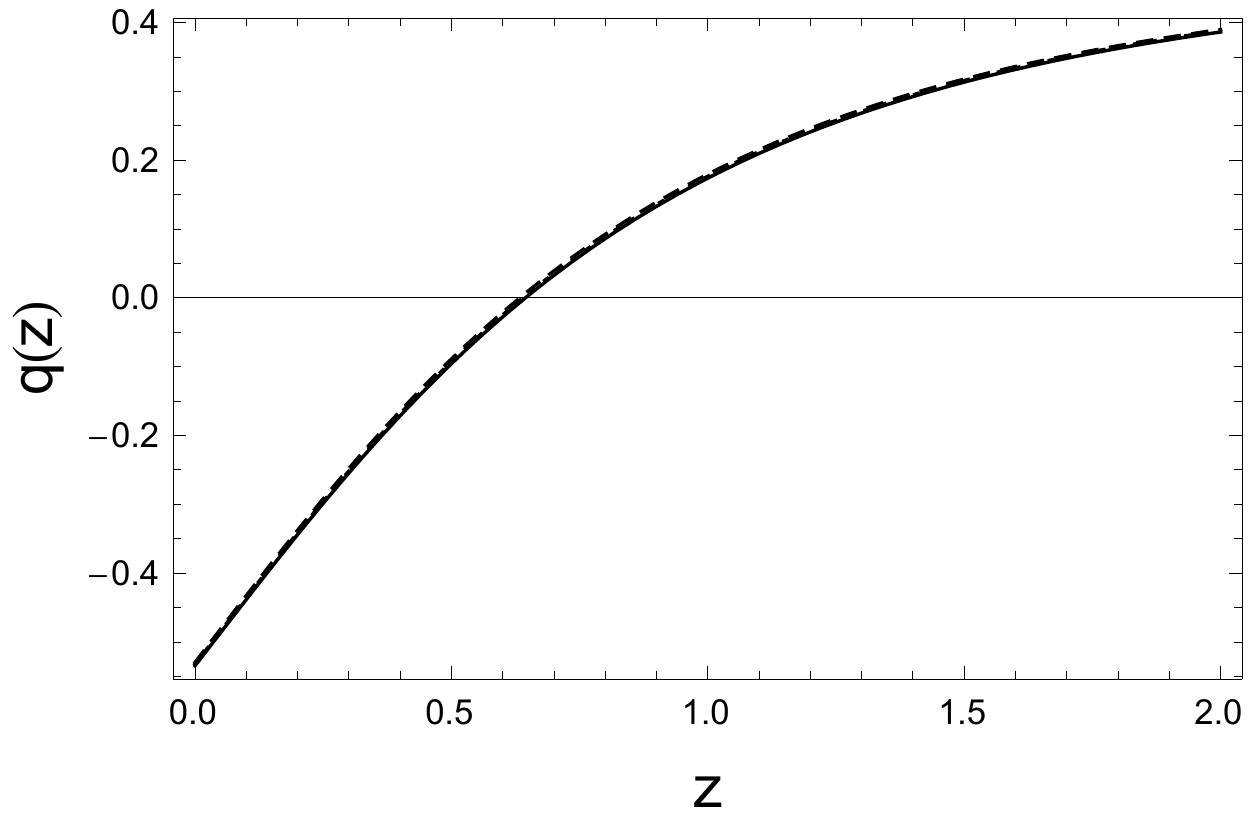}~~~
  \includegraphics[width=0.45\textwidth]{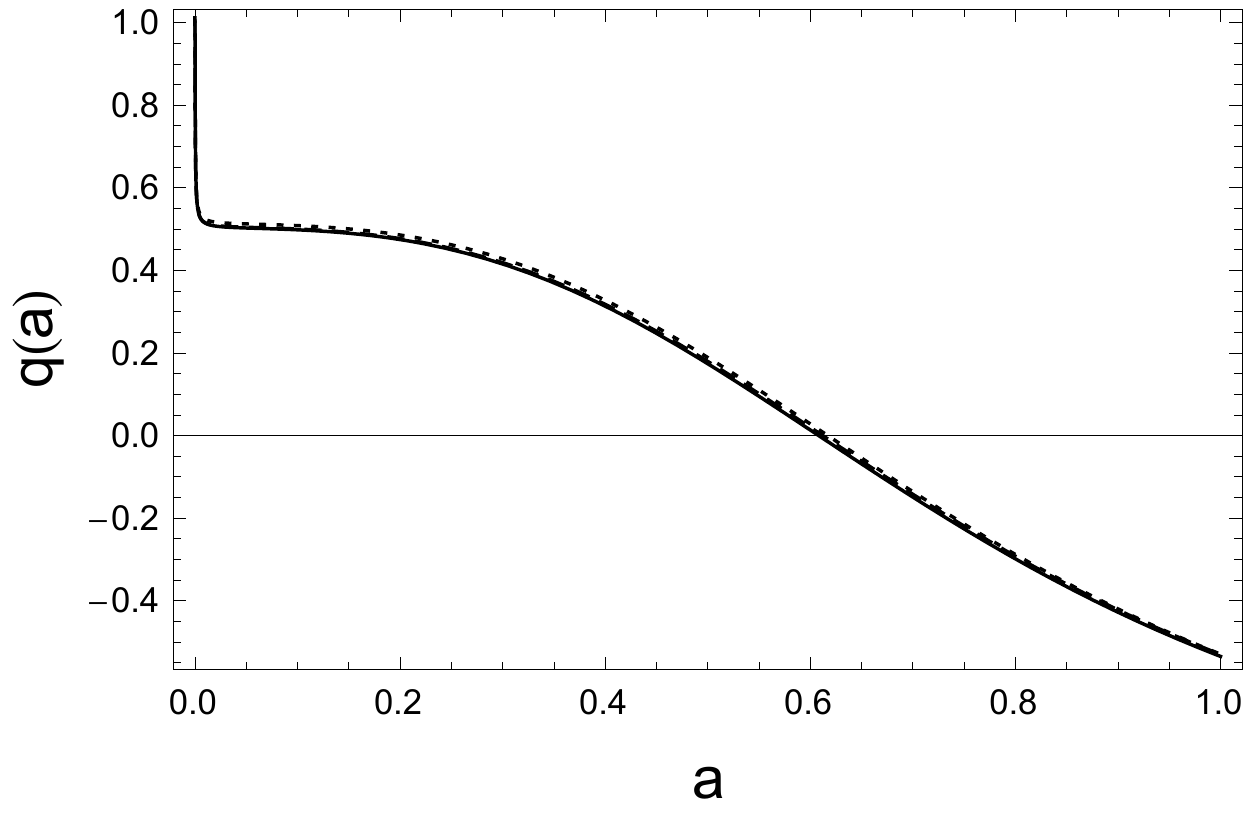}
\caption{Transition from deceleration to acceleration for the model (\ref{h}). Solid line is for standard $\Lambda$CDM model; dashed line for varying--$G$--entropic--$\Lambda$CDM model; dotted line is for varying--$c$--entropic--$\Lambda$CDM model. The values of the parameters are taken from Table I and from the \textit{Planck} data ($\Lambda$CDM).} \label{fig1}
\end{minipage}
\end{figure*}

\section{Data Analysis}
\label{data}

The analysis has involved the largest updated set of cosmological data available so far, and includes: Type Ia Supernovae (SNeIa); Baryon Acoustic Oscillations (BAO); Cosmic Microwave Background (CMB); and a prior on the Hubble constant parameter, $H_{0}$.

\subsection{Type Ia Supernovae}

We used the SNeIa (Supernovae Type Ia) data from the JLA (Joint-Light-curve Analysis) compilation \citep{JLA}. This set is made of $740$ SNeIa obtained by the SDSS-II (Sloan Digital Sky Survey) and SNLS (Supenovae Legacy Survey) collaboration, covering a redshift range $0.01<z<1.39$. The $\chi^2_{SN}$ in this case is defined as
\begin{equation}
\chi^2_{SN} = \Delta \boldsymbol{\mathcal{F}}^{SN} \; \cdot \; \mathbf{C}^{-1}_{SN} \; \cdot \; \Delta  \boldsymbol{\mathcal{F}}^{SN} \; ,
\end{equation}
with $\Delta\boldsymbol{\mathcal{F}}^{SN} = \mathcal{F}^{SN}_{theo} - \mathcal{F}^{SN}_{obs}$, the difference between the observed and the theoretical value of the observable quantity $ \mathcal{F}^{SN}$; and $\mathbf{C}_{SN}$ the total covariance matrix (for a discussion about all the terms involved in its derivation, see \citep{JLA}). For JLA, the observed quantity will be the predicted distance modulus of the SNeIa, $\mu$, given the cosmological model and two other quantities,  the stretch (a measure of the shape of the SNeIa light-curve) and the color. It will read
\begin{equation}\label{eq:m_jla}
\mu(\boldsymbol{\theta}) = 5 \log_{10} [ D_{L}(z, \boldsymbol{\theta_{c}}) ] - \alpha X_{1} + \beta \mathcal{C} + \mathcal{M}_{B} \; ,
\end{equation}
where $D_{L}$ is the luminosity distance
\begin{equation}\label{eq:dL}
D_{L}(z, \boldsymbol{\theta_{c}} )  = \frac{c_{0}}{H_{0}} (1+z) \ \int_{0}^{z} \frac{\mathrm{d}z'}{E(z',\boldsymbol{\theta_{c}})} \; ,
\end{equation}
with $H(z) \equiv H_{0} E(z)$ (following \citep{JLA}, we assume $H_{0} = 70$ km/s Mpc$^{-1}$), $c_{0}$ the speed of light here and now, and $\boldsymbol{\theta_{c}}$ the vector of cosmological parameters. The total vector $\boldsymbol{\theta}$ will include $\boldsymbol{\theta_{c}}$ and the other fitting parameters, which in this case are: $\alpha$ and $\beta$, which characterize the stretch-luminosity and color-luminosity relationships; and the nuisance parameter $\mathcal{M}_{B}$, expressed as a step function of two more parameters, $\mathcal{M}^{1}_{B}$ and $\Delta_{m}$:
\begin{equation}
\mathcal{M}_{B} = \begin{cases} \mathcal{M}^{1}_{B} & \mbox{if} \quad M_{stellar} < 10^{10} M_{\odot}, \\
\mathcal{M}^{1}_{B} + \Delta_{m} & \mbox{otherwise}.
\end{cases}
\end{equation}
Further details about this choice are given in Ref. \citep{JLA}. The formula (\ref{eq:mu_jla}) stands for the constant $c$ cases; when $c$ is varying according to (\ref{ansaetze}), it is modified into \cite{PLB14,JCAP14}
\begin{equation}\label{eq:mu_jla}
D_{L}(z, \boldsymbol{\theta_{c}} )  = \frac{c_{0}}{H_{0}} (1+z) \ \int_{0}^{z} \frac{(1+z')^{-n}}{E(z',\boldsymbol{\theta_{c}})} \mathrm{d}z'\; .
\end{equation}

\subsection{Baryon Acoustic Oscillations}

The $\chi^2_{BAO}$ for Baryon Acoustic Oscillations (BAO) is defined as
\begin{equation}
\chi^2_{BAO} = \Delta \boldsymbol{\mathcal{F}}^{BAO} \; \cdot \; \mathbf{C}^{-1}_{BAO} \; \cdot \; \Delta  \boldsymbol{\mathcal{F}}^{BAO} \; ,
\end{equation}
where the quantity $\mathcal{F}^{BAO}$ can be different depending on the considered survey. We used data from the WiggleZ Dark Energy Survey \citep{WiggleZ_0}, evaluated at redshifts $z=\{0.44,0.6,0.73\}$, and given in Table~1 of \citep{WiggleZ}; in this case the quantities to be considered are the acoustic parameter
\begin{equation}\label{eq:AWiggle}
A(z, \boldsymbol{\theta_{c}}) \equiv 100  \sqrt{\Omega_{m} \, h^2} \frac{D_{V}(z,\boldsymbol{\theta_{c}})}{c_{0} \, z} \, ,
\end{equation}
and the Alcock-Paczynski distortion parameter
\begin{equation}\label{eq:FWiggle}
F(z, \boldsymbol{\theta_{c}}) \equiv (1+z)  \frac{D_{A}(z,\boldsymbol{\theta_{c}})\, H(z,\boldsymbol{\theta_{c}})}{c_{0}} \, ,
\end{equation}
where, $D_{A}$ is the angular diameter distance
\begin{equation}\label{eq:dA}
D_{A}(z, \boldsymbol{\theta_{c}} )  = \frac{c_{0}}{H_{0}} \frac{1}{1+z} \ \int_{0}^{z} \frac{\mathrm{d}z'}{E(z',\boldsymbol{\theta_{c}})} \; ,
\end{equation}
and $D_{V}$ is a combination of the physical angular-diameter distance $D_A$ (tangential separation) and Hubble parameter $H(z)$ (radial separation) defined as
\begin{equation}\label{eq:dV}
D_{V}(z, \boldsymbol{\theta_{c}} )  = \left[ (1+z)^2 D^{2}_{A}(z,\boldsymbol{\theta_{c}}) \frac{c_{0} \, z}{H(z,\boldsymbol{\theta_{c}})}\right]^{1/3}.
\end{equation}
When dealing with varying $c$, Eqs.~(\ref{eq:AWiggle})~(\ref{eq:FWiggle})~(\ref{eq:dA})~(\ref{eq:dV}) have to be changed into \cite{PLB14,JCAP14}:
\begin{equation}\label{eq:AWiggle2}
A(z, \boldsymbol{\theta_{c}}) \equiv 100  \sqrt{\Omega_{m} \, h^2} \frac{D_{V}(z,\boldsymbol{\theta_{c}})}{c_{0} (1+z)^{-n} \, z} \, ,
\end{equation}
\begin{equation}\label{eq:FWiggle2}
F(z, \boldsymbol{\theta_{c}}) \equiv (1+z)  \frac{D_{A}(z,\boldsymbol{\theta_{c}})\, H(z,\boldsymbol{\theta_{c}})}{c_{0}(1+z)^{-n}} \, ,
\end{equation}
\begin{equation}\label{eq:dA2}
D_{A}(z, \boldsymbol{\theta_{c}} )  = \frac{c_{0}}{H_{0}} \frac{1}{1+z} \ \int_{0}^{z} \frac{(1+z')^{-n}}{E(z',\boldsymbol{\theta_{c}})} \mathrm{d}z'\; ,
\end{equation}
\begin{equation}\label{eq:dV2}
D_{V}(z, \boldsymbol{\theta_{c}} )  = \left[ (1+z)^2 D^{2}_{A}(z,\boldsymbol{\theta_{c}}) \frac{c_{0} (1+z)^{-n} \, z}{H(z,\boldsymbol{\theta_{c}})}\right]^{1/3}.
\end{equation}
We have also considered the data from SDSS-III Baryon Oscillation Spectroscopic Survey (BOSS) DR$10$-$11$, described in \citep{BOSS1,BOSS2}. Data are expressed as
\begin{equation}
D_{V}(z=0.32) = (1264 \pm 25) \frac{r_{s}(z_{d})}{r^{fid}_{s}(z_{d})} \, ,
\end{equation}
and
\begin{equation}
D_{V}(z=0.57) = (2056 \pm 20) \frac{r_{s}(z_{d})}{r^{fid}_{s}(z_{d})} \, ,
\end{equation}
where $r_{s}(z_{d})$ is the sound horizon evaluated at the dragging redshift $z_{d}$, and $r^{fid}_{s}(z_{d})$ is the same sound horizon but calculated for a given fiducial cosmological model used, being equal to $149.28$ Mpc \citep{BOSS1,BOSS2}. The redshift of the drag epoch is well approximated by \citep{Eisenstein}
\begin{equation}\label{eq:zdrag}
z_{d} = \frac{1291 (\Omega_{m} \, h^2)^{0.251}}{1+0.659(\Omega_{m} \, h^2)^{0.828}} \left[ 1+ b_{1} (\Omega_{b} \, h^2)^{b2}\right]
\end{equation}
where
\begin{eqnarray}
b_{1} &=& 0.313 (\Omega_{m} \, h^2)^{-0.419} \left[ 1+0.607 (\Omega_{m} \, h^2)^{0.6748}\right], \nonumber \\
b_{2} &=& 0.238 (\Omega_{m} \, h^2)^{0.223}.
\end{eqnarray}
And the sound horizon is defined as:
\begin{equation}\label{eq:soundhor}
r_{s}(z) = \int^{\infty}_{z} \frac{c_{s}(z')}{H(z',\boldsymbol{\theta_{c}})} \mathrm{d}z'\, ,
\end{equation}
with the sound speed
\begin{equation}\label{eq:soundspeed}
c_{s}(z) = \frac{c_{0}}{\sqrt{3(1+\overline{R}_{b}\, (1+z)^{-1})}} \,
\end{equation}
and
\begin{equation}
\overline{R}_{b} = 31500 \Omega_{b} \, h^{2} \left( T_{CMB}/ 2.7 \right)^{-4}\, ,
\end{equation}
with $T_{CMB} = 2.726$ K.

We have also added data points from Quasar-Lyman $\alpha$ Forest from SDSS-III BOSS DR$11$ \citep{Lyman}:
\begin{eqnarray}
\frac{D_{A}(z=2.36)}{r_{s}(z_{d})} &=& 10.8 \pm 0.4 \nonumber \\
\frac{c_{0}}{H(z=2.36) r_{s}(z_{d})}  &=& 9.0 \pm 0.3.
\end{eqnarray}
When working with varying $c$ models, of course, we will have to change $D_{A}$ and $D_{V}$ as described above, and also the sound horizon, through the definition of the sound speed, Eq.~(\ref{eq:soundspeed}), which now will be \cite{PLB14,JCAP14}
\begin{equation}
c_{s}(z) = \frac{c_{0} (1+z)^{-n}}{\sqrt{3(1+\overline{R}_{b}\, (1+z)^{-1})}}.
\end{equation}
Thus, we will have three different contributions to $\chi^{2}_{BAO}$, e.g., $\chi^{2}_{WiggleZ},\chi^{2}_{BOSS},\chi^{2}_{Lyman}$, depending on the data sets we consider.

\subsection{Cosmic Microwave Background}

The $\chi^2_{CMB}$ for Cosmic Microwave Background (CMB) is defined as
\begin{equation}
\chi^2_{CMB} = \Delta \boldsymbol{\mathcal{F}}^{CMB} \; \cdot \; \mathbf{C}^{-1}_{CMB} \; \cdot \; \Delta  \boldsymbol{\mathcal{F}}^{CMB} \; ,
\end{equation}
where $\mathcal{F}^{CMB}$ is a vector of quantities taken from \citep{WangWang}, where \textit{Planck} first data release is analyzed in order to give a set of quantities which efficiently summarize the information contained in the full power spectrum (at least, for the cosmological background), and can thus be used in alternative to the latter \citep{WangMukherjee}. The quantities are the CMB shift parameters:
\begin{eqnarray}
R(\boldsymbol{\theta_{c}}) &\equiv& \sqrt{\Omega_m H^2_{0}} \frac{r(z_{\ast},\boldsymbol{\theta_{c}})}{c_{0}} \nonumber \\
l_{a}(\boldsymbol{\theta_{c}}) &\equiv& \pi \frac{r(z_{\ast},\boldsymbol{\theta_{c}})}{r_{s}(z_{\ast},\boldsymbol{\theta_{c}})}\, ,
\end{eqnarray}
and the baryonic density parameter, $\Omega_b \, h^{2}$. Again, $r_{s}$ is the comoving sound horizon, but evaluated at the photon-decoupling redshift $z_{\ast}$, given by the fitting formula \citep{Hu}:
\begin{equation}{\label{eq:zdecoupl}}
z_{\ast} = 1048 \left[ 1 + 0.00124 (\Omega_{b} h^{2})^{-0.738}\right] \left(1+g_{1} (\Omega_{m} h^{2})^{g_{2}} \right) \, ,
\end{equation}
with
\begin{eqnarray}
g_{1} &=& \frac{0.0783 (\Omega_{b} h^{2})^{-0.238}}{1+39.5(\Omega_{b} h^{2})^{-0.763}} \nonumber \\
g_{2} &=& \frac{0.560}{1+21.1(\Omega_{b} h^{2})^{1.81}} \, ;
\end{eqnarray}
while $r$ is the comoving distance defined as:
\begin{equation}
r(z, \boldsymbol{\theta_{c}} )  = \frac{c_{0}}{H_{0}} \int_{0}^{z} \frac{\mathrm{d}z'}{E(z',\boldsymbol{\theta_{c}})} \mathrm{d}z'\; .
\end{equation}
When considering varying $c$ models, again, the sound horizon will change as described above, and the comoving distance will be \cite{PLB14,JCAP14}
\begin{equation}
r(z, \boldsymbol{\theta_{c}} )  = \frac{c_{0}}{H_{0}} \int_{0}^{z} \frac{(1+z')^{-n}}{E(z',\boldsymbol{\theta_{c}})} \mathrm{d}z'\; ,
\end{equation}
and the shift parameter $R$ will become
\begin{equation}
R(\boldsymbol{\theta_{c}}) \equiv \sqrt{\Omega_m H^2_{0}} \frac{r(z_{\ast},\boldsymbol{\theta_{c}})}{c_{0} (1+z_{\ast})^{-n}} .
\end{equation}

Moreover, we have added a gaussian prior on the Hubble constant, $H_{0}$
\begin{equation}
\chi^{2}_{H_{0}} = \frac{(H_{0}- 69.6)^{2}}{0.07^2}
\end{equation}
derived from \citep{Bennett}.

Thus, the total $\chi^2_{Tot}$ will be the sum of: $\chi^{2}_{SN},\chi^{2}_{WiggleZ},\chi^{2}_{BOSS},\chi^{2}_{Lyman},\chi^{2}_{CMB},\chi^{2}_{H_{0}}$. We minimize $\chi^2_{Tot}$ using the Markov Chain Monte Carlo (MCMC) method.

Finally, we should make a few comments about the parameters which will be constrained. The total parameters vector $\boldsymbol{\theta_{c}}$ will be equal to $\{\Omega_{m}, \Omega_{b}, h, q, \gamma, \alpha\, \beta\, M^{1}_{B}, \Delta_{M}\}$ when considering the varying $G$ cases, and $\{\Omega_{m}, \Omega_{b}, h, n, \gamma, \alpha\, \beta\, M^{1}_{B}, \Delta_{M}\}$ when considering the varying $c$ ones. The actual observationally fitted components of this vector are given in Table \ref{tab:results}.

The parameter $h$ is defined in a standard way by $H_{0} \equiv 100 \, h$. The density parameters entering $H(z)$ are $\Omega_{m},\Omega_{r},\Omega_{v}$; assuming zero spatial curvature, we can express $\Omega_{v} = 1-\gamma - \Omega_{m}-\Omega_{r}$, in order to ensure the condition $E(z=0)=1$. Moreover, the radiation density parameter $\Omega_{r}$ will be defined \citep{WMAP} as the sum of photons and relativistic neutrinos
\begin{equation}
\Omega_{r} = \Omega_{\gamma} (1+0.2271 \mathcal{N}_{eff})\, ,
\end{equation}
where $\Omega_{\gamma} = 2.469 \times 10^{-5} \, h^{-2}$ for $T_{CMB}= 2.726$ K; and the number of relativistic neutrinos is assumed to be $\mathcal{N}_{eff} = 3.046$.

\subsection{Results}

\begin{figure*}[h]
\centering
\includegraphics[width=8.cm]{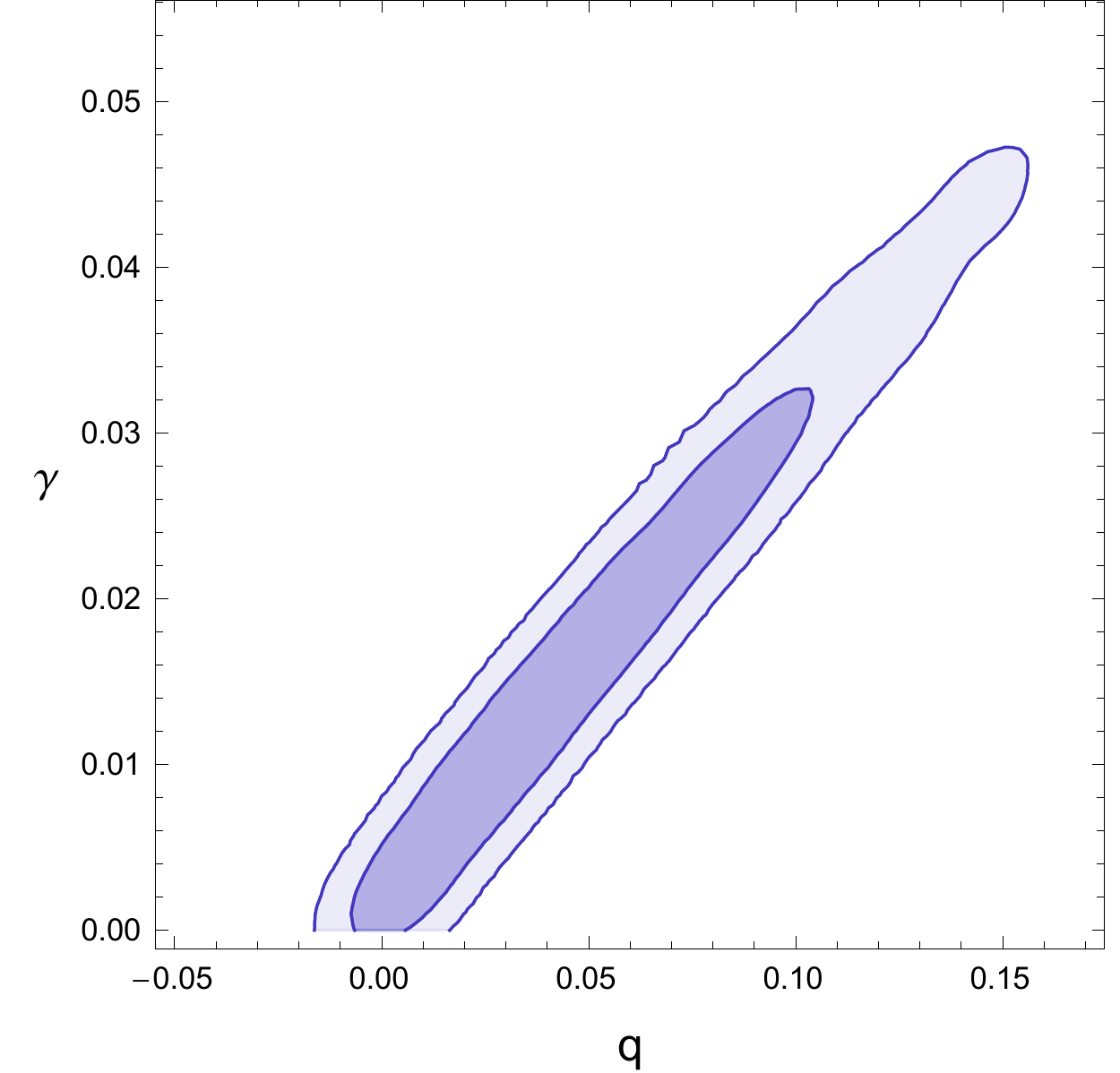}~~~
\includegraphics[width=8.cm]{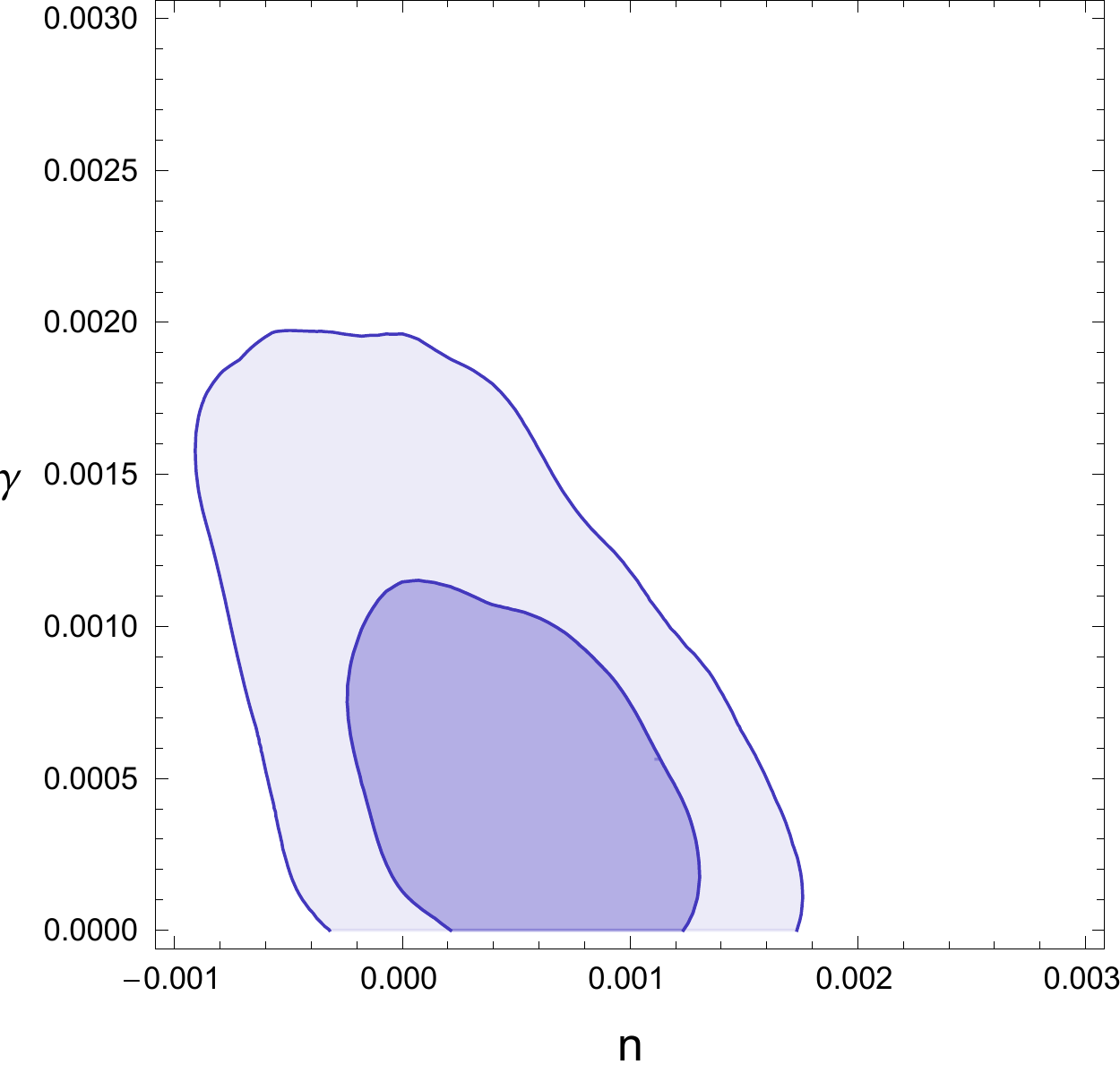}
\caption{\textit{(Left.)} Varying $G$ scenario: $68\%$ and $95\%$ confidence levels for $q$ and $\gamma$. \textit{(Right.)} Varying $c$ scenario: $68\%$ and $95\%$ confidence levels for $n$ and $\gamma$.}\label{fig:contours}
\end{figure*}

Our main result is presented in Fig. \ref{fig:contours}. First novelty is that we have found the observational bounds on the Hawking temperature coefficient $\gamma$ which (on the theoretical basis) was usually taken to of order of unity $O(1)$. Our evaluation gives that it should be of the order of $10^{-2} - 10^{-4}$. This difference is not unexpected, because the $O(1)$ estimation was based on purely theoretical considerations, with no previous connection to data. Now, we  show that observations are not consistent with such large values of $\gamma$. Instead, it is at least two orders of magnitude less. Thus, the entropic force in the model we have considered gives only a small contribution. Similar results were obtained in \citep{Basilakos09,Gomez141}. Another novelty is the bound on the variability of the speed of light $c$ and the gravitational constant $G$. According to them, in the entropic scenario we have investigated both $G$ (Fig. \ref{fig:contours}, left panel) and $c$ (Fig. \ref{fig:contours}, right panel) should be increasing with the evolution of the universe. Bearing in mind that the speed of light is related to the inverse of the fine structure constant defined as
\be
\label{fine}
\alpha = \frac{e^2}{\hbar c} ,
\ee
where $e$ is the electron charge and $\hbar$ is the Planck constant, by using (\ref{ansaetze}) and (\ref{fine}) one has
\be
\frac{\Delta c}{c} = - \frac{\Delta \alpha}{\alpha} = n \frac{\Delta a}{a} \sim \frac{n}{10},
\ee
then one can derive from Table \ref{tab:results} and Fig. \ref{fig:contours} that the change in $c$ and so in $\alpha\; (\Delta \alpha / \alpha_{0})$ from our fit is $\sim 10^{-5}$ in a redshift range $[1;2]$ ($n = 4.9 \cdot 10^{-4} > 0$), while other observational bounds, in the same range (see table II of Ref. \cite{carlos} which is based on \cite{LP1,alphaMolaro,alphaChand,alphaAgafonova}), give $\Delta \alpha / \alpha_{0} \sim 10^{-6}$. But still our estimation is compatible with other cosmological constraints, as the ones derived from CMB \textit{Planck} first release, see \cite{OBrian}. Moreover, recent observations show that both positive and negative values of $n$ are possible (the so-called $\alpha-$dipole \cite{webb}).

Finally, we can enumerate some general conclusions as follows:
\begin{itemize}
  \item the entropic scenario plus varying $c$ and/or $G$ is quite indistinguishable from a \textit{pure-$\Lambda$CDM model}, that is why we call it an  \textit{entropic-$\Lambda$CDM model}. Present data is still unable to differentiate between the two scenarios;
  \item the model obtained ({\it entropic-$\Lambda$CDM cosmology}) is a variation of the exchange of energy between vacuum and matter model studied in Refs. \cite{Basilakos09,Grande11,Gomez141}.
  \item the best fit for the value of the Hawking temperature coefficient $\gamma$ is quite different from the theoretical values used in literature, i.e. $\gamma=3/(2\pi)$ or $1/2$; it should be pointed out that other considered entropic scenarios have the values of $O(1)$ (e.g. \cite{i1});
  \item the model with small values of the parameter $\gamma$ is equivalent to a dynamical vacuum model with small variation of the vacuum energy studied in Refs. \cite{Basilakos09,Gomez141};
  \item the value for $\gamma$ is compatible with zero since we were able to put only an upper limit to it. This would mean that the Hawking temperature were zero for the models under study;
  \item it is also clear that we still have the deceleration-acceleration transition, as we show in the plot of the relation for $q(z)$ and also for $q(a)$ in Fig. \ref{fig1}, where our models are compared with a standard $\Lambda$CDM resulting, as said above, barely distinguishable.
\end{itemize}

The models we have studied here involve a mixture of matter and the dark energy fluid which is typically the energy of vacuum with small modifications due to the variability of $c$ and $G$. This means that the discussion of the structure formation problem (perturbation equations, the formation of the structures, linear growth rate) is similar to those of dynamical vacuum models given in Ref. \cite{Grande11} with $\gamma$ parameter here being analogous to $\nu$ parameter of that reference. In fact, the models the models $III$ and $IV$ of Ref. \cite{Grande11} are indistinguishable from $\Lambda$CDM while the models $I$ and $II$ exhibit some difference what can be seen from Fig.~1 of \cite{Grande11}, where the density contrast and the linear growth rate of clustering are shown.

{\renewcommand{\tabcolsep}{2.mm}
{\renewcommand{\arraystretch}{1.25}
\begin{table*}[h]
\begin{minipage}{\textwidth}
\caption{Observational parameters of the entropic models under study.}\label{tab:results}
\centering
\resizebox*{\textwidth}{!}{
\begin{tabular}{|c|ccccccccc|}
\hline
$id.$ & $\Omega_{m}$ & $\Omega_{b}$ & $h$ & $q/n$ & $\gamma$ & $\alpha$ & $\beta$ & $\mathcal{M}^{1}_{B}$ & $\Delta_{m}$ \\
\hline \hline
$G = G_{0}\; a^{q}$ & $0.314^{+0.009}_{-0.008}$ & $0.0453^{+0.0009}_{-0.0009}$ & $0.698^{+0.007}_{-0.007}$ & $0.048^{+0.042}_{-0.033}$ & $<0.022$ & $0.141^{+0.007}_{-0.006}$ & $3.106^{+0.077}_{-0.087}$ & $-19.044^{+0.018}_{-0.019}$ & $-0.071^{+0.023}_{-0.023}$ \\
\hline
$c = c_{0}\; a^{n}$ & $0.311^{+0.007}_{-0.007}$ & $0.046^{+0.001}_{-0.001}$ & $0.696^{+0.007}_{-0.007}$ & $0.00049^{+0.00049}_{-0.00053}$ & $<0.0007$ & $0.141^{+0.007}_{-0.007}$ & $3.100^{+0.080}_{-0.080}$ & $-19.043^{+0.018}_{-0.018}$ & $-0.070^{+0.023}_{-0.022}$ \\
\hline \hline
\end{tabular}}
\end{minipage}
\end{table*}}}

\section{Conclusions}
\label{conclusions}
In this paper we extended the entropic cosmology onto the framework of the theories with varying gravitational constant $G$ and varying speed of light $c$. We discussed the consequences of such variability onto the entropic force terms and the boundary terms using three different approaches which possibly relate thermodynamics, cosmological horizons and gravity. We started with a general set of the field equations which described varying constants entropic cosmology with a general form of the entropic terms. In the first approach we derived the continuity equation from the first law of thermodynamics, Bekenstein entropy as well as Hawking temperature to fit the general entropic terms to this continuity equation. We found appropriate single-fluid accelerating cosmological solutions to these field equations. We also discussed the constraints on the models which come from the second law of thermodynamics. In the second approach we derived the entropic force for varying constants, defined the entropic pressure, and finally  modified the continuity and the acceleration equations. Then, we determined the Friedmann equation and gave single-fluid accelerating cosmological solutions as well. Finally, in the third approach we got gravitational Einstein field equations using the heat flow through the horizon to which Bekenstein entropy and Hawking temperature were assigned.

 We have also examined some of the many-fluid (first accelerating and then decelerating) entropic models against observational data (supernovae, BAO, and CMB). We have used data from JLA compilation of SDSS-II and SNLS collboration (supernovae), WiggleZ Dark Energy Survey and SDSS-III Baryon Oscillation Spectroscopic Survey (BOSS) as well as \textit{Planck} data (CMB).
 We found that the observational bound on the Hawking temperature coefficient $\gamma$ was much smaller ($10^{-2} - 10^{-4}$) than it is usually assumed on the theoretical basis to be of order of unity $O(1)$. We have also found that in our entropic models $G$ should be diminishing while $c$ should be increasing with the evolution of the universe. Our bound on the variation of $c$ being $\Delta c/ c \sim 10^{-5}$ is at least one order of magnitude weaker than observational bound obtained from analysis of the quasar spectra.

\section{Acknowledgements}

This project was financed by the Polish National Science Center Grant DEC-2012/06/A/ST2/00395.



\end{document}